\date{}
\documentclass[11]{article}

\usepackage[top=2.5cm,left=2.5cm,right=2.5cm,bottom=2.5cm]{geometry}

\usepackage{amsmath,epic,eepic}
\usepackage{epsf,psfig}
\usepackage{graphicx}
\usepackage{amssymb}
\usepackage{color}

\newcommand{\ignore}[1]{}

\newcommand{\be}{\begin{equation}}
\newcommand{\ee}{\end{equation}}

\newcommand{\yn}[1]{\textcolor{green}{[YN: #1]}}

\begin{document}
\title{Empirical Limitations on\\ High Frequency Trading Profitability}
\author{Michael Kearns,
Alex Kulesza,
Yuriy Nevmyvaka\\
Computer and Information Science\\
University of Pennsylvania}
\date{}

\maketitle

\begin{abstract}
  Addressing the ongoing examination of high-frequency trading
  practices in financial markets, we report the results of an
  extensive empirical study estimating the maximum possible
  profitability of the most aggressive such practices, and arrive at
  figures that are surprisingly modest.  By ``aggressive'' we mean any
  trading strategy exclusively employing market orders and relatively
  short holding periods.  Our findings highlight the tension between
  execution costs and trading horizon confronted by high-frequency
  traders, and provide a controlled and large-scale empirical
  perspective on the high-frequency debate that has heretofore been
  absent. Our study employs a number of novel empirical methods,
  including the simulation of an ``omniscient'' high-frequency trader
  who can see the future and act accordingly.
\end{abstract}


\section{Introduction}

The recent financial crisis has been accompanied by rising popular,
media, and regulatory alarm over what is broadly called high frequency
trading (HFT in the sequel). The overarching fear is that quantitative
trading groups, armed with advanced networking and computing
technology and expertise, are in some way victimizing retail traders
and other less sophisticated parties.  
The HFT debate often conflates distinct phenomena, confusing, for
instance, dark pools and flash trading, which are essentially new
market mechanisms, with HFT itself, which is a type of trading
behavior applicable to both existing and emerging exchanges.  The core
concern regarding HFT, however, is relatively straightforward: that
the ability to electronically execute trades on extraordinarily short
time scales,\footnote{Often measured in milliseconds or less, and
  aided by colocation, the placement of trading servers directly at
  the exchange.}  combined with the quantitative modeling of massive
stores of historical data, permits a variety of practices unavailable
to most parties.  A broad example would be the discovery of very
short-term informational advantages (for instance, by detecting large,
slow trades in the market) and profiting from them by trading rapidly
and aggressively.

Despite the growing controversy over HFT,\footnote{As of this writing,
  SEC and congressional investigations into HFT practices were
  ongoing.} there appear to be no objective, large-scale empirical
studies of the potential profitability and impact of HFT.  The purpose
of this paper is to provide one such study.  Our main conclusion is
perhaps unexpected: for at least one broad class of ``aggressive''
HFT, the total available market size --- that is, the maximum profit
that could conceivably be realized using this type of HFT --- is
rather modest (relative to the size of the market and the
publicly-reported trading revenues of other market participants).
More precisely, we demonstrate an upper bound of \$21 billion for the
{\em entire universe of U.S. equities in 2008\/} at the longest
holding periods, down to \$21 million or less for the shortest holding
periods (see discussion of holding periods below).  Furthermore, we
believe these numbers to be vast overestimates of the profits that
could actually be achieved in the real world.  These figures should be
contrasted with the approximately \$50 {\em trillion\/} annual trading
volume in the same markets.  We believe our findings are of interest
in their own right as well as potentially relevant to the ongoing
debate over HFT.

In our study, we make a crucial distinction between {\em passive\/}
HFT, in which a HFT strategy exclusively places limit orders that are
not immediately marketable --- and thus acts as a provider of
liquidity to the market --- and {\em aggressive\/} HFT, in which only
market orders are used, and thus the HFT must pay the attendant
execution costs of crossing the bid-ask spread. In this study we focus
{\em only on aggressive HFT\/}, and we shall argue that this is the
variety of HFT that should be the primary focus of any concern, since
the presence of passive HFT can only provide price and liquidity
improvement to any trading counterparties.\footnote{In practice, HFT
  strategies may employ mixtures of limit and market orders.  We
  expect to extend our methodology to passive HFT shortly.}

For aggressive HFT, there is a fundamental tension between two
basic quantities: the {\em horizon\/} or {\em holding period\/}, as measured
by the length of time for which a (long or short) position in a stock
is held; and the {\em costs\/} of trading, as measured by (at least)
the bid-ask spread that must be crossed by market orders on entry and
liquidation of the position. In order for a trade to be profitable,
the position must be held long enough for favorable price movement
sufficient to overcome the trading costs. The shorter the holding
period, the more extreme (and thus less frequent) the relative price
movements must be for profitability.  As we shall argue, rational
concern over HFT should focus on short holding periods --- measured in
seconds or less --- since at longer holding periods, the advantages of
rapid exchange access and low latency are obviously diminished compared
to the general trading population.  While this tension between horizon
and costs is well-understood in quantitative finance, it has not
before been empirically studied on a large scale in the context of
HFT, which is our main contribution.

At the core of our experimental study is a novel but simple technique
we call the {\em Omniscient Trader Methodology\/} (or OTM). Given the
constraints of aggressive HFT, and armed with two large and rich
historical trading data sets, we compute the profit or loss of all
possible trades available to the HF trader {\em in hindsight\/}, and
reach empirical overestimates of profitability by counting {\em only
the profitable trades.\/} In this way we deliberately remove the
greatest difficulty in real quantitative trading --- that of
predicting which trades will be profitable --- and obtain what are
certainly gross overestimates of the profits realizable in practice to
a trader engaging in aggressive HFT. Of course, such a study is
interesting only if these overestimates are still surprisingly modest,
which is exactly the conclusion we shall establish.

The overview of our methodology is as follows. We begin with the
highest-resolution data available: every message from the NASDAQ stock
exchange, which permits exact replication of the full historical
evolution of the order books for any chosen stock on that exchange. We
use this data to simulate the aforementioned OTM; however, the sheer
volume of messages and computational intensity of these experiments
preclude running them on all of the thousands of stocks traded on
NASDAQ. Furthermore, NASDAQ is only one of the exchanges on which US
stocks are traded, so does not provide the full picture of potential
profits.  To remedy these shortcomings, we employ a two-step process:
first we use the OTM to upper bound the profitability of HFT on a
small set of the most liquid (and therefore most profitable) stocks on
NASDAQ, and then use a slightly less detailed data set and regression
methods to scale up our estimates to a much larger universe of all US
stocks, and across all exchanges.  At every step of this process we
are careful to err on the side of overestimation in order to ensure
that our final figures are upper bounds on total HFT profitability.

Our primary contributions can be summarized as follows:
\begin{itemize}
\item We provide the first controlled and objective empirical study of
  the profitability of (aggressive) HFT based on extensive historical
  market microstructure data, and discuss its implications.
\item We develop new empirical methods for the estimation of the
  profitability of various classes of trading strategies, including
  the Omniscient Trader Methodology, and techniques for extrapolating
  profitability from a smaller to a larger set of stocks and
  exchanges.
\item We report detailed findings relating a variety of fundamental
  market microstructure variables, including price movement, spreads,
  trading volume and profitability.
\end{itemize}

\section{Related Work}
\nobreak

A significant motivation for our work has been the recent attention
paid to HFT in the mainstream and online media.  Some major news
outlets and industry experts have expressed concern over the rising
popularity of HFT and its influence on the price formation process
\cite{Duhigg09,Moyer09,Lash09,Goldstein09}, while others have
concluded that HFT is unobjectionable or even beneficial
\cite{Levitt09,Heires09,Schack09}. The debate centers on a basic
difference in perspective: Does HFT allow large firms to make money by
preying on other market participants with less sophisticated
technology and slower access to exchanges?  Or do high-frequency
traders effectively compete amongst themselves to provide the service
of enhanced liquidity to everyone else?  Evidence presented by both
sides tends to be largely anecdotal, in part because of the opaque
nature of financial firms, particularly those practicing HFT.
Academic research in this area is also limited, probably for similar
reasons.

Nonetheless, there has been at least one attempt to study HFT
empirically. Aldridge \cite{Aldridge09} applies an omniscient
methodology related to ours, but examines foreign exchange
trading. She reports per-period returns of 0.04\% - 0.06\% at short
trading intervals (which are comparable to the returns found in our
experiments), but is silent on total {\em profits\/}, which is our
focus.  Furthermore, her analysis centers around the Sharpe ratio, a
measure of return per unit of risk.  She shows that the simulated
Sharpe ratio can rise above 5,000 at 10 second trading intervals, and
in turn labels HFT strategies as extremely safe, but it is unclear
what these numbers mean given that her simulation assumes an
omniscient trader who by definition is never at risk of losing money.
Sharpe ratios can obviously be driven arbitrarily high if we have
perfect knowledge; here we demonstrate that profits cannot.

Other studies have focused on the qualitative effects of HFT.  Chaboud
et al. \cite{Chaboud09} study foreign exchange markets, and find that
HFT does not cause an increase in volatility.  Furthermore, they show
that computerized HF traders provide liquidity more often than human
traders following information shocks.  Hendershott et al. analyze
electronic trading of US equities \cite{Hendershott10}, and find that
HFT improves liquidity on NYSE in large-cap stocks.  The same authors
examine German markets in \cite{Hendershott09}, where they conclude
that HFT provides liquidity when it is expensive and takes it when it
is cheap.

Others have tried to estimate the total (actual) profits due to HFT.
Tabb et al. arrive at a number of \$21 billion or more \cite{Iati09},
but a second report from the same group gives a figure of \$8.5
billion \cite{Tabb09}.  Donefer proposes \$15--25 billion in
\cite{Donefer09}.  Schack and Gawronski of Rosenblatt Securities claim
that all of these numbers are too high \cite{Schack09}, but do not
themselves offer a specific number.  Goldman Sachs, considered by some
to be a leader in HFT, claims in a note to clients that less than 1\%
of their revenues come from HFT \cite{Goldman09}, which bounds the
annual HFT profits of Goldman Sachs at \$0.5 billion. All of these
sources ostensibly account for both passive and aggressive HFT, but
Arnuk and Saluzzi address aggressive HFT specifically and put it
at \$1.5--3 billion per year \cite{Arnuk09}.  

In contrast to these
studies, which are based largely on proprietary data and involve
subjective estimates of realized profits, we propose a verifiable,
empirical bound on the {\em maximum feasible\/} HFT profits using
publicly available data.  We also focus specifically on aggressive
trading, which as we will discuss is the form of HFT with the greatest
potential for harm.

\section{Microstructure Preliminaries}
\label{sec:pre}

We now provide some necessary background on the fundamental mechanism
underlying the majority of modern electronic exchanges, including
those of the U.S. equities markets.

The {\em open limit order book\/} is a market mechanism for
implementing a type of continuous double auction.  Suppose we wish to
purchase 1000 shares of Microsoft (MSFT) stock on such an exchange.
To do so, we submit a {\em limit order\/} that specifies not only the
desired volume of 1000 shares, but also the maximum price we are
willing to pay.  Assuming that nobody is currently willing to sell at
the price we have requested, our order gets placed in the {\em buy
  order book\/}, a list of all current offers to buy MSFT shares,
sorted by price, with the highest price (known as the bid) at the top
of the book.  If there are multiple limit orders at the same price,
they are ordered by time of arrival, with older orders getting
priority.  Symmetrically, the exchange maintains a sell order book
containing offers to sell MSFT shares, sorted with the lowest price
(known as the ask) at the top.  Thus, the tops of the books always
consist of the most competitive offers.  The bid and ask prices
together are referred to as the inside market, and the difference
between them is known as the spread.

\begin{figure}[t]
\centering \includegraphics[height=3.8in]{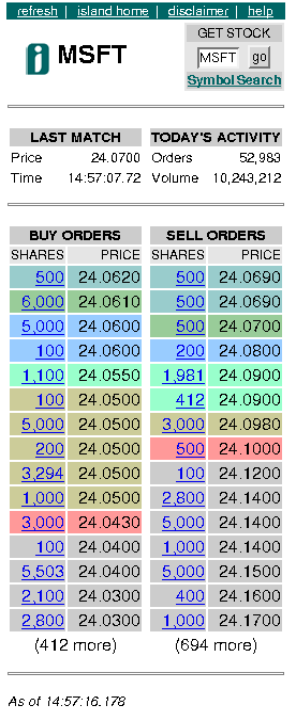}
\caption[]{\small Sample order books for MSFT.} \vspace{-.2in}
\label{fig:islbook}
\end{figure}

Figure~\ref{fig:islbook} depicts an actual snapshot of an order book
for MSFT\footnote{Note that both the figure and accompanying
numerical example are outdated in the sense that 
NASDAQ now enforces decimalization, but the mechanism is unchanged.}.
The bid is \$24.062, and the ask is \$24.069.  The spread
is \$0.007.  If we were to submit a buy order at \$24.04, our order
would be placed immediately after the extant order for 5,503 shares at
\$24.04.

If a buy (respectively, sell) limit order comes in above the ask (below
the bid), a trade will occur.  The incoming order will be matched with
orders on the opposing book, beginning at the top, until either the
incoming order's volume is filled or no further matching is
possible. For example, suppose in Figure~\ref{fig:islbook} a buy order
for 2000 shares arrives with a limit price of \$24.08. This order will
be partially filled by the two 500-share sell orders at \$24.069, the
500-share order at \$24.07, and the 200-share order at \$24.08, for a
total of 1700 shares executed. The remaining 300 shares of the
incoming order would become the new top of the buy book; the new bid
would be \$24.08 and the new ask would be \$24.09.  Note that
executions happen at the prices of the {\em resting limit orders\/},
not at the price of the incoming order.  Thus, consuming orders deeper
in the opposing book yields progressively worse prices for the party
initiating execution.

Any order can be canceled at any time prior to execution. Full 
exchange rules are often more complicated, including hidden orders,
trading halts, crosses, inter-exchange linkages, and other special
mechanisms.  Nonetheless, the vast majority of trading in modern
markets takes place via the simple mechanism described above, and this 
is exactly the mechanism we replicate in our simulations.

\section{Constraints on HFT}

In order to empirically estimate the potential profitability of HFT in
a meaningful way, we need to commit to a precise definition.  Given
our intention to apply omniscience in order to upper bound
profitability, we need constraints that disallow arbitrarily large,
long-term trades that could yield essentially unbounded profits in
hindsight.  The main constraints we shall impose are those of
aggressive order placement and short holding periods.

\subsection{Aggressive Order Placement}

Traders in modern electronic exchanges have the choice between
``passive'' or ``aggressive'' order placement: they can either place
limit orders that do not instigate any immediate executions and lie in
their respective books awaiting possible later execution; or they can
place immediately marketable orders that cross the spread, eat into
the opposing book, and pay both the spread costs and potentially
higher costs for ``deeper'' shares. With regards to the debate
surrounding HFT, {\em we propose that aggressive orders are the
greatest cause for any concerns about the negative impacts on trading
counterparties.\/} The reason for this is both simple and
well-understood in finance: passive order placement can only {\em
improve\/} the market for any counterparties, both in prices and
volumes.  By placing a passive limit order, a trader can only reduce
spreads and provide more shares and available price levels for the
market. Indeed, this is why many exchanges actually give {\em
rebates\/} for orders that lie in the books but are eventually
executed --- they are providing liquidity --- whereas fees are
routinely charged for aggressive orders, which are removing
liquidity. Furthermore, if one of the advantages of (and concerns
over) HFT is the ability to very rapidly take and liquidate positions
to profit from short-term informational advantages, aggressive order
placement is necessary: if we have a predictive advantage for (say) 1
second, we can't realize it by waiting for the other side of the
market to come to us --- we must initiate the entry and exit trades.

For these reasons, in the current study we shall restrict our
attention to aggressive order placement. A concrete example of a type
of HFT we are excluding by this choice is market-making, in which a
trader tries to perpetually maintain both buy and sell passive limit
orders, profiting whenever pairs of such orders are executed without
ever acquiring significant (long or short) inventory.  While
market-making is a natural and common type of trading strategy, we
again note that it is {\em not\/} generally cited as part of the
concern over HFT.

\subsection{Short Holding Periods}

As discussed in the introduction, the part of the definition of HFT
that most parties seem to be able to agree on is the ability to place
(and therefore execute or cancel) orders on an extremely short time
scale. However, if such an ability is used to instigate positions that
are then {\em held\/} for hours or even many seconds, the advantage of
HFT is already lost, since nowadays even the slowest electronic access
to the exchanges would typically be measured in hundreds of
milliseconds.  In other words, either (much of) the profitability of a
given trade is realized almost immediately after it is executed, in
which case only a fast trader can capture the value, or the
profitability is realized over a longer horizon, in which case almost
any competent electronic trading platform can capture it.  Also,
holding positions for longer periods of time is associated with higher
risk, whereas HFT is often portrayed as a lower risk,
technology-fueled activity that hinges on fast access to the
marketplace to take advantage of fleeting opportunities.

We thus posit that with regards to the debate surrounding HFT, {\em
only trades held for short periods should be considered.\/} In our
study we shall consider holding periods as short as 10 milliseconds
and as long as 10 seconds; in our view the latter already stretches
the definition of HFT, but we include it to err on the side of
generality.

\section{Methodology and Data}

At the core of our empirical approach is the {\em Omniscient Trader
  Methodology\/} (OTM). While there are technical details to be
discussed shortly, the underlying idea behind the OTM is quite simple:
given complete historical data on a given stock, we can identify
exactly those trades that {\em would\/} have been profitable {\em in
  hindsight\/} --- that is, given complete knowledge of the trading
future of the stock in question. We thus simulate an Omniscient Trader
(OT), whose profitability is obviously an upper bound on the
profitability of {\em any\/} realistic strategy that must make on-line
trading decisions based only on the past, and not future data.

Of course, to be meaningful and interesting the OTM must be applied to
some reasonably restricted class of trades, otherwise trades such as
``buy all available shares of Google at its IPO and hold them until
the present'' will produce wild profitability but little
insight. Since our primary interest is in HFT, as discussed above the
main constraints we shall place on the OT are those of short holding
periods and aggressive trade execution.

We shall apply the OTM to two different sources of raw trading
data. The first source consists of complete event information (order
placements, cancellations, modifications, and trade executions) for a
set of 19 highly liquid NASDAQ stocks; this data is sufficient to
fully reconstruct the complete historical order books for each stock
at any moment in 2008.  (The data is limited to 19 stocks because of
the computational intensity of the resulting experiments, discussed
below.)  We shall refer to this data set as the {\em order book\/}
data.  The second data source (known in finance as Trade and Quote, or
TAQ, data) consists of a somewhat cruder summary of liquidity, but for
a much larger set of stocks and exchanges.  Both of our data sets are
commercially available and widely used in quantitative finance.

Our overall methodology is to first apply the OTM to the order book
data, which provides us with a very detailed view of profitability and
the relationships between a number of fundamental microstructure
variables for the limited set of 19 stocks on NASDAQ. We then use
contemporaneous TAQ data and regression methods on the same 19 stocks
in order to construct reliable models for scaling up our estimates to
the full universe of U.S. equities and exchanges (including NYSE).
Here we will first describe the OTM as applied to the order book data;
in Section~\ref{sec:taq} we describe the use of TAQ data for the
broader universe.

We have implemented a rather powerful and flexible software platform
that fully reconstructs the historical order books for a given NASDAQ
stock at any chosen time resolution, and can simulate (or ``backtest''
in the parlance of finance) a wide variety of quantitative trading
strategies. In this framework, we simulate the profitability of the
following OT, whose only parameter is the {\em holding period\/} $h$:
\begin{itemize}
\item At each time $t$, the OT may either buy or sell $v$ shares, for
  every integer $v \geq 0$.  The purchase or sale of the $v$ shares
  occurs at market prices; thus according to the standard
  U.S. equities limit order mechanism, the OT crosses the spread
  and consumes the first $v$ shares on the opposing order book,
  receiving possibly progressively worse prices for successive shares.
\item If at time $t$ the OT bought/sold $v$ shares, at time $t+h$ it
  must liquidate this position and sell/buy the shares back, again by
  crossing the spread and paying market prices on the opposing book.
  (In Section~\ref{sec:perspective} we discuss the effects of allowing
  a variable holding period.)
\item At each time $t$, the OT makes only that trade (buying or
  selling, and the choice of $v$) that {\em optimizes
    profitability\/}. Obviously it is this aspect of the OT that
  requires knowledge of the future. Note that if no trade at $t$ has
  positive profitability, the OT does nothing.
\end{itemize}

This design for the OT accurately captures the fundamental tension of
(aggressive) HFT.  If we denote the bid-ask spread at time $t$ by
$s_t$, it is clear that the purchase or sale of even a single share by
the OT will incur transaction costs on the order of at least
$\frac{1}{2}(s_t + s_{t+h})$ --- the mean of the spreads at the onset
and liquidation of the position. Larger values for $v$ will increase
these costs, since eating further into the opposing books effectively
widens the spreads. Thus, in order for a trade of $v$ shares to be
profitable, the share price {\em must have time to change enough to
cover the spread-based transaction costs\/}.  (Since we omnisciently
optimize between buying and selling, as well as the trade volume,
sufficient movement either up or down will result in some profitable
trade.)  Of course, the smaller the holding period $h$, the less
frequently such fluctuations will occur --- indeed, we find that for
sufficiently small $h$, the vast majority of the time the optimal
choice is $v=0$ --- that is, no trade is made by the OT.

We assert that this tension between the depth of liquidity (as
represented by the spreads and volumes available at different prices
in the order books) and holding periods sufficiently long to permit
profitable price movement is the fundamental source of limitation of
aggressive HFT profitability.  One of our primary contributions is to
carefully measure this tension experimentally, and show that the
limitations are indeed severe at short holding periods (i.e., true
HFT).

We now briefly remark on a number of further details on our data and
experimental methodology.

\begin{itemize}
\item The order book data provides, for every NASDAQ stock, every
  single message sent to and from the exchange in 2008 --- order
  placements, modification, cancellations, and executions --- with
  high-resolution (sub-millisecond) timestamps. This data is
  voluminous: for a liquid stock such as AAPL, the data for just a
  single day averages 138 MB, and for efficiency it must be
  decompressed and recompressed online rather than in batch.
\item The data volume and computational intensity of our experiments
  preclude examining the entire NASDAQ universe (and as we shall see,
  this is wholly unnecessary for our purposes since we employ methods
  for estimating profits on a much broader universe). We thus selected
  a set of 19 stocks on which to conduct our initial experiments.
  These stocks were sampled from among the most liquid NASDAQ stocks,
  since as we shall see liquidity is highly correlated with
  profitability of HFT.\footnote{The ticker symbols of the specific
    stocks examined are AAPL, ADBE, AMGN, AMZN, BIIB, CELG, COST,
    CSCO, DELL, EBAY, ESRX, GILD, GOOG, INTC, MSFT, ORCL, QCOM, SCHW,
    and YHOO.  } Even restricted to these 19 stocks, a run of our OTM
  experiments for just a single holding period consumes many CPU-days.
\item In our OTM simulations, the books are used to compute the most
  profitable trade at each moment, but immediately ``reset'' to their
  previous state following any action by the OT.  In other words, we
  assume the trades of the OT have {\em no impact\/} on the market
  (not even reducing the liquidity available to the OT in the future),
  since we cannot realistically propagate such effects from historical
  data.  Thus we err strongly on the side of overestimating
  profitability; in real markets trades nearly always have negative
  price impacts for the trader.
\item We discretize time into distinct instances at which the HFT is
  able to consider placing a profitable trade under the OTM, since
  otherwise, having no lasting effect on the market, the OT could make
  the same profitable trade an arbitrary number of times in rapid
  succession.  At the extreme, we might allow new trades to be placed
  any time the books change, but this leads to massive overcounting
  --- for instance, a cancellation of an order deep in the books does
  not really provide a ``new'' trading opportunity from a picosecond
  before the cancellation.  Thus, for the results reported here, we
  chose to allow trading every 10 milliseconds conditioned on there
  being {\em any\/} change --- in prices, volumes, numbers of orders,
  etc. --- at the bid or ask prices.  Again, it is nearly certain that
  we are overcounting profitability (i.e., counting what is
  essentially the same profitable trade many times) due to small,
  inconsequential changes to the books \footnote{The main reason that
    we do not systematically address this overcounting --- e.g.  by
    removing orders against which we have already executed --- is that
    the methodology described preserves the original price formation
    dynamics. Otherwise, we would have to impose some model of order
    book evolution following our hypothetical trades.}.  In any case,
  we conducted the same experiments using a variety of other choices,
  including the logical extreme of every exchange event, and the
  findings were qualitatively the same.
\item HFT was widespread during 2008, and the order book data we use already 
reflects that HF activity.  Thus there might be concern that we are only
measuring the {\em additional\/} profits not already extracted by real HFT
during that year. This is not true, however, since 
the OTM captures {\em all\/} profitable trades, including those that were 
realized by actual HF traders in 2008.  This is because the OT is 
infinitely fast and can immediately capitalize on every opportunity, 
regardless of whether a real trader did the same moments later.
More precisely, if there was a profitable trade actually realized by a HFT party
and thus reflected in our data, the OTM will capture this same trade by
making it an instant earlier in our simulation.

\end{itemize}

In short, using the OTM we err on the side of optimism and
overestimation of HFT profitability in as many dimensions as possible:
complete knowledge of the future, computation of the optimally
profitable volume to trade, rapid instance generation leading to
overcounting of profitable trades, exclusion of exchange fees and
commissions, and so on.  Still, we shall see that powerful insights
and limitations can be gleaned from our results.  Of course, there
are also limitations to our approach, which we discuss in
Section~\ref{sec:perspective}.

\section{Omniscient Order Book Trading}

\label{sec:qat}

\begin{figure*}[htbp]
\centering
\includegraphics[width=6.5in]{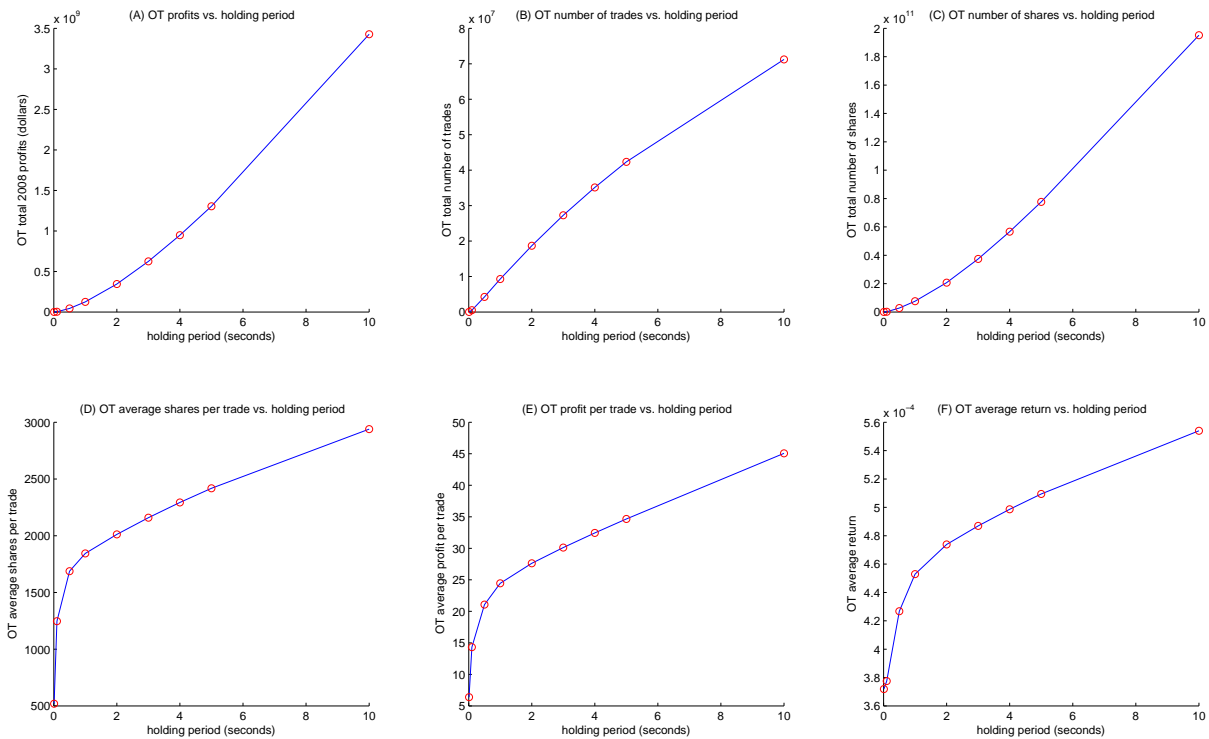}
\caption[]{\small OT profits and other quantities vs. holding period
  for the 19 NASDAQ stocks on the order book data.}
\label{fig:holding}
\end{figure*}

Our first set of results is presented in the six panels of
Figure~\ref{fig:holding}. In each panel, the x-axis measures the
holding period, sampled at values of 0.01, 0.1, 0.5, 1, 2, 3, 4, 5,
and 10 seconds.  The y-axes measure a variety of different quantities
against these holding periods.

Panel (A) of Figure~\ref{fig:holding} presents what is perhaps our
simplest and most fundamental finding.  For each of the holding
periods, it plots the total 2008 OT profits (in dollars) for the 19
stocks in our order book data set. Because of the nature of the OT, we
can interpret these values as the {\em maximum possible total
profitability for aggressive HFT in these stocks in all of 2008.\/}
The most striking aspects of this plot are the absolute numbers
themselves: even at the longest holding period of 10 seconds (which
again stretches even the most liberal interpretation of ``high
frequency'') the total 2008 OT profits are only \$3.4
billion --- a figure any individual person or trading group would
be happy to reap, but perhaps small considering the omniscience assumption
and high liquidity of these stocks.
Furthermore, profitability
at shorter holding periods falls off rapidly, down to just 
\$62,000 for the shortest (10ms) holding period. We conclude that the
combination of aggressive order placement with short holding periods
{\em severely\/} limits the potential profitability of HFT in these 19
stocks; we shall see shortly that this story remains essentially
unchanged when we scale up to a much larger universe of stocks and
exchanges.

Panel (A) establishes that OT profitability decays strongly with
shorter holding periods, but does not quantify the cause of this
decay; for instance, it could be that there are fewer profitable
trades present at shorter holding periods, or that the trades present
at shorter holding are less profitable on average, or have lower
returns. In panels (B) through (F) of Figure~\ref{fig:holding} we
examine these and other quantities as a function of holding period in
our experiments.

Panels (B) and (C) examine two measures of total OT trading volume ---
the total number of profitable trades executed, and the total number
of shares traded in those profitable trades. We see that these plots
show a decay at short holding periods similar to that for total
profitability. For instance, the ratio of total profitability at 1
second holding to that at 10 second holding is 0.036; the analogous
figures for number of trades and total number of shares are 0.13 and
0.039. So profitability is falling roughly in tandem with the number
of trading opportunities and (especially) the total share volume.

Panel (D) plots the average number of shares per trade against holding
period.  Remembering that the OTM explicitly optimizes the number of
shares to maximize the total profitability of each trade, we can view
this plot as measuring the depth of the opposing book that can be
profitably consumed at each holding period. As expected, this decays
with shorter holding periods --- the trading costs of eating deep into
the opposing books are not overcome by favorable price movement at
shorter horizons --- but we see strong sublinearity to this plot, with
the ratio of shares per trade at 1 second to 10 seconds being 0.63 ---
qualitatively larger than for the measures above. Only at the very
shortest holding periods do we see a precipitous decline in the
optimal trade volume.\footnote{We note that the fact that several
  thousand shares per trade are executed at the longest holding period
  also suggests that our profit overestimation is likely greatest
  there, since such large trades would almost certainly have
  significant negative longer-term price impacts.}  A similar picture
holds for the average dollar profit per trade (panel (E)) and average
returns (panel (F)) --- they are relatively steady in the 1 to 10
second range, but fall quickly at the shortest (sub-second) holding
periods.

In short, everything gets worse for the OT at short holding periods:
profits plummet, and the first-order explanation for this is simply
that there are many fewer profitable trading opportunities as a result
of the trade-off between horizon and execution costs. In addition,
when profitable trades are present, they are of smaller volume and
lower returns.

\begin{figure*}[htbp]
\centering
\includegraphics[width=6.5in]{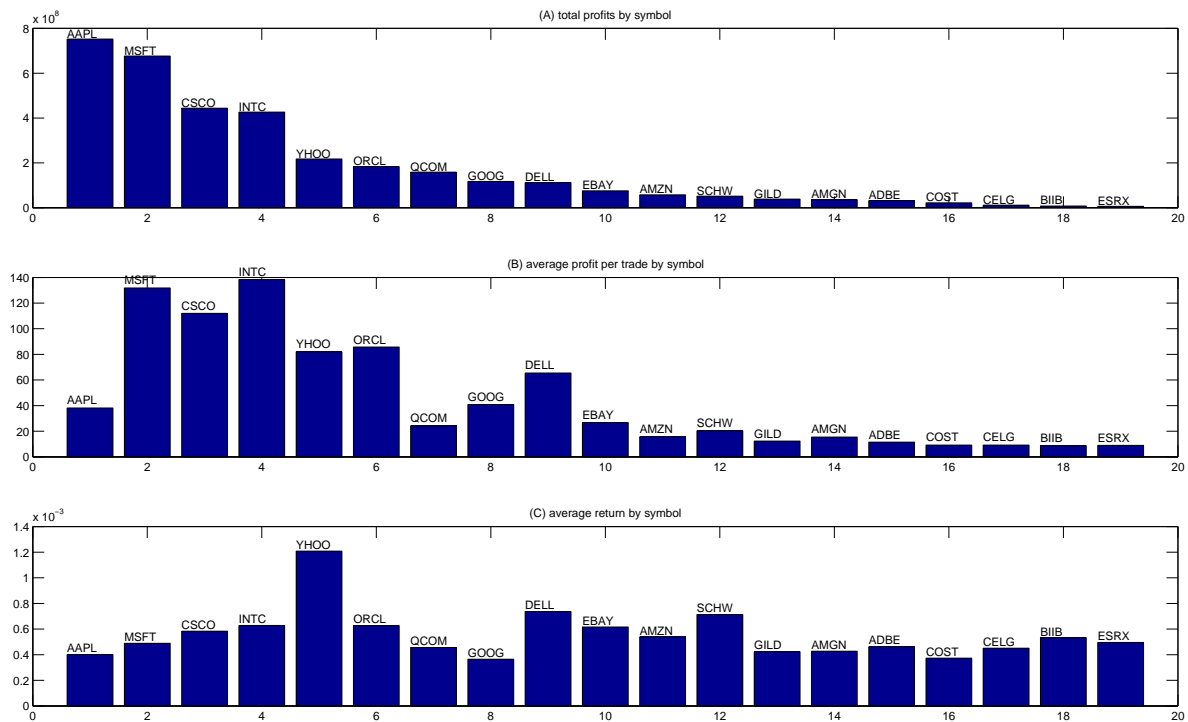}
\caption[]{\small OT profits and returns by stock on the order book data.} 
\label{fig:names}
\end{figure*}

The plots of Figure~\ref{fig:holding} aggregate quantities across all
19 of the stocks in our order book data set; however, also of interest
is how these quantities are distributed across the individual stocks.
For simplicity we focus on just the longest (and most profitable)
holding period of 10 seconds, but the story remains qualitatively
similar at shorter holding.

The bar charts of Figure~\ref{fig:names} show total 2008 OT profits
(panel (A)), average profit per trade (panel (B)), and average return
(panel (C)) on a stock-by-stock basis, with each bar labeled by the
relevant ticker symbol for the stock in question.  The stocks are
sorted in each plot by their total profitability.

Perhaps most striking is the rapid decline in profitability as we move
to less profitable stocks (panel (A)).  Indeed, just 5 of the 19
stocks --- AAPL, MSFT, CSCO, INTC and YHOO --- already account for
73\% of the total profits, while the worst 10 account for only 13\%.
Thus, not only is aggressive HFT surprisingly modest in profitability,
that profitability is also highly concentrated in just a handful of
the most liquid stocks.  (Recall that high liquidity was a primary
rationale for our choice of these 19 stocks in the first place, and we
shall see shortly that liquidity is highly correlated with
profitability.)

Panels (B) and (C) of Figure~\ref{fig:names} demonstrate that while
the average profit per trade is again rather concentrated among a
handful of the most liquid stocks, returns remain relatively constant
across stocks. OT profitability is primarily driven by liquidity, and
there are a handful of stocks that are outliers in this regard,
providing much greater usable depth of book to the OT. Returns,
however --- which do not reward the higher absolute profits that come
with greater trade volume --- are driven by price volatility, and
there are no stocks here with volatility orders of magnitude greater
than the average.

\begin{figure}[htbp]
\centering \includegraphics[width=3.0in]{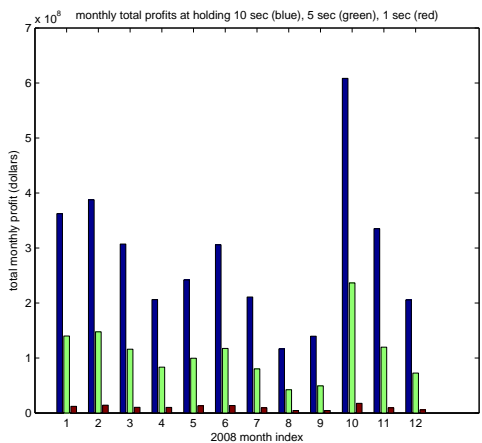}
\caption[]{\small Profits by month and holding period} 
\label{fig:monthly}
\end{figure}

We conclude our experimental OT order book data results with a brief
temporal study, presented in Figure~\ref{fig:monthly}, which shows the
total 19-stock OT profits in 2008 on a monthly basis for holding
periods of 10, 5, and 1 seconds. Of particular note is the spike in
profitability in October 2008, which corresponds to the peak of the
financial crisis, shortly after the collapse of Lehman Brothers and
the chain of events it set in motion. The fact that certain HFT groups
enjoyed great profitability during the crisis, as many banks and hedge
funds were forced to liquidate assets, is well-known and indeed part
of the controversy surrounding HFT. Interestingly, this spike
diminishes as one moves to shorter holding periods --- the ratio of
October to average monthly profits decreases monotonically from 2.13
at 10 seconds holding to 1.64 at 10 ms holding (not plotted). Again,
we have a horizon/cost trade-off at work here, as it is also
well-known that spreads widened considerably during the crisis, making
it harder to ``victimize'' liquidating parties at the shortest time
scales under aggressive execution methods.

\section{Market-Wide Extrapolation}
\label{sec:taq}

The results described so far suffer from two primary limitations: they
consider only a set of 19 NASDAQ-listed stocks (albeit highly liquid
ones); and they consider only trading activity within the actual
NASDAQ exchange (the so-called ``primary'' exchange for these stocks),
whereas trading occurs in a broader set of venues (``composite''
activity). In this section we address these limitations and
extrapolate our results to all stocks and all exchanges by utilizing a
broader but somewhat less detailed data source known as Trade and
Quote (TAQ).  These extrapolations lead to our final estimated upper
bound on the profits available to HFT in 2008 on the entirety of the
US stock market.  First, we estimate a primary-to-composite conversion
ratio (in a given name, around 50\% of profits come from the primary
exchange); then we propose a simple model that relates HFT
profitability to the number of quote updates in a given stock, and use
it to get market-wide results.

\subsection{Exchanges and TAQ}

Our full order book data are collected solely from the NASDAQ
exchange; however, in reality the 19 stocks, though traded primarily
on NASDAQ, are also traded on a variety of alternative exchanges that
in some instances may offer better pricing.  In order to account for
the additional profits available to traders operating across multiple
exchanges, we rely on the fact that commercially available TAQ data
are sufficient to reconstruct the bid and ask prices, as well as the
total number of shares available at those prices, at any past moment,
both on the primary exchange (NASDAQ for our 19 stocks) as well as a
composite of all US exchanges.  By comparing the two scenarios, we can
estimate the additional profits available on secondary exchanges.

\begin{figure}[tbp]
\centering \includegraphics[width=2.8in]{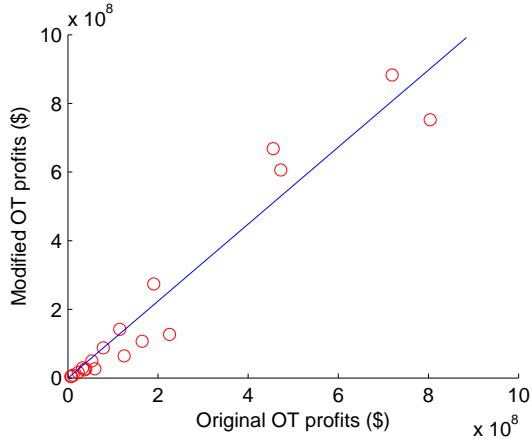}
\caption[]{\small Modified vs original OT profit estimates} 
\label{fig:taqqat}
\end{figure}

\begin{figure*}[tbp]
\centering \includegraphics[width=6.5in]{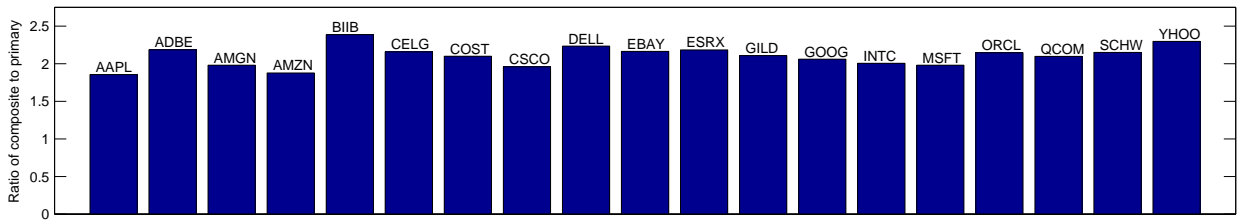}
\caption[]{\small Composite vs. primary exchange profits} 
\label{fig:primcomp}
\end{figure*}

Unfortunately, TAQ data are limited in that they do not record
liquidity offered beyond the current bid and ask prices.  As a result,
we cannot fully simulate our omniscient trader, which utilizes the
full depth of the books.  We therefore proceed by first establishing
an empirical correspondence between the OT discussed so far and a
modified OT that we can simulate using TAQ data alone.  We then use
the modified OT to extrapolate our results.

Recall that our OT to this point has been allowed to trade any number
of shares at each time $t$ and must liquidate the position at time
$t+h$, where $h$ is the holding period.  Since TAQ data do not allow
us to compute the profitability of trades that reach beyond the top of
the book, we modify the OT as follows.  At each time $t$, the OT can
buy or sell $v$ shares, where $v \geq 0$ is an integer bounded by the
number of shares available to buy (sell) at the ask (bid) at time $t$
as well as the number of shares available to sell (buy) at the bid
(ask) at time $t+h$.  For example, if 100 shares are offered at the
ask price at time $t$ and 50 shares are offered at the bid price at
time $t+h$, then the OT can only consider buying up to 50 shares at
time $t$.  (Since the trader is omniscient, it can compute these
limits even though they depend on future information.)  As before, any
position taken at time $t$ must be liquidated at time $t+h$.

We began by computing the theoretical profits of the modified OT from
TAQ data using primary (NASDAQ-only) pricing, closely replicating the
setting from Section~\ref{sec:qat}.  Figure~\ref{fig:taqqat} compares
these profits with those of the original OT for the 19 stocks across
all of 2008.  The holding period for both traders is 10 seconds,
though the results do not differ significantly at shorter holding
periods.  Note that due to small differences between the two data
sources, the modified OT (simulated on TAQ data) occasionally achieves
higher profits than the original OT (simulated on full order book
data).  Nonetheless, the correlation coefficient between the two
traders' profits is 0.969.  Because of this close linear relationship,
we proceed under the assumption that the original OT's profits are
closely estimated as a multiple of the modified OT's profits.  This
allows us to directly apply extrapolation ratios computed using TAQ
data to the profits reported in Section~\ref{sec:qat}.

We next used TAQ data to compute the profitability of the modified OT
given the more inclusive composite pricing.  Because spreads can only
shrink as additional exchanges are allowed to compete, we expect to
see greater profits from aggressive HFT in this setting.
Figure~\ref{fig:primcomp} shows the ratio of the modified OT's profits
under composite pricing to those under primary pricing for each of the
19 stocks at the most profitable holding period of 10 seconds.  On
average, profits increase by a factor of 2.1; this ratio is nearly the
same across stocks, suggesting that the inclusion of secondary
exchanges has a consistent, measurable effect on potential HFT
profits.

Finally, we estimated composite profits of the original OT by
multiplying the profits reported in Section~\ref{sec:qat} by the
corresponding ratios depicted in
Figure~\ref{fig:primcomp}.\footnote{We also apply a small correction
term to account for several days in 2008 where our order book data is
incomplete for techncial reasons.  This increases our final results by
less than 5\%.}  The results, shown in Table~\ref{tab:composite}, are
estimated upper bounds on the total profits available to the OT with
10 second holding and access to all US exchanges for all of 2008.

\begin{table}[htbp]
\centering \begin{tabular}{cr|cr}
AAPL	&\$1,490 M & ESRX	&\$14 M\\
ADBE	&\$71 M    & GILD	&\$84 M\\
AMGN	&\$73 M    & GOOG	&\$257 M\\
AMZN	&\$112 M   & INTC	&\$913 M\\
BIIB	&\$19 M    & MSFT	&\$1,424 M\\
CELG	&\$26 M    & ORCL	&\$409 M\\
COST	&\$48 M    & QCOM	&\$346 M\\
CSCO	&\$924 M   & SCHW	&\$114 M\\
DELL	&\$257 M   & YHOO	&\$518 M\\
EBAY	&\$170 M   & &
\end{tabular}
\caption[]{\small Composite profit upper bounds for the original 
                  OT at 10 second holding, in millions}
\label{tab:composite}
\end{table}

\subsection{Regression}

Having estimated bounds on the maximum profits available from
high-frequency trading for 19 high-volume NASDAQ stocks, we proceed to
extrapolate these figures to a wider universe of stocks.  While we
could use TAQ data to simulate the modified OT for many stocks, the
costs of simulation make it impractical to do so.  We therefore turn
to a simple regression approach, estimating potential profits as a
function of high-level information extracted from the TAQ data.  One
could imagine using a large and diverse array of statistics as the
basis for a very accurate profit model; however, complex approaches
run the risk of overfitting and may not generalize well to other
stocks.  Instead, we found that a single simple measurement is
sufficient to closely predict the simulated profits.  Specifically,
for each stock we compute the number of times during 2008 that the
best available offer changed, either in price or quantity, on the
stock's primary exchange.  This value is simply the total number of
quotes reported in the TAQ data.

\begin{figure}[tbp]
\centering
\includegraphics[width=2.8in]{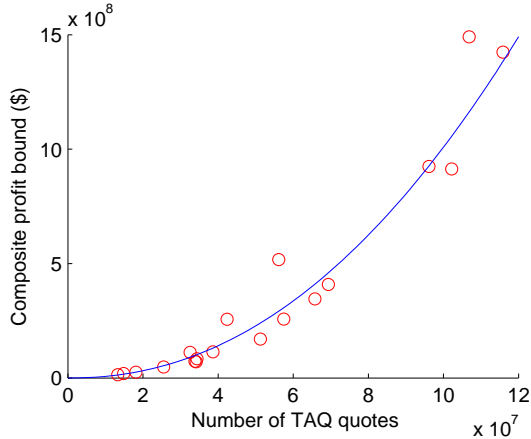}
\caption[]{\small Composite profit upper bounds (10 second holding) vs
  number of quotes}
\label{fig:regs}
\end{figure}

\begin{figure}[tbp]
\centering \includegraphics[width=2.8in]{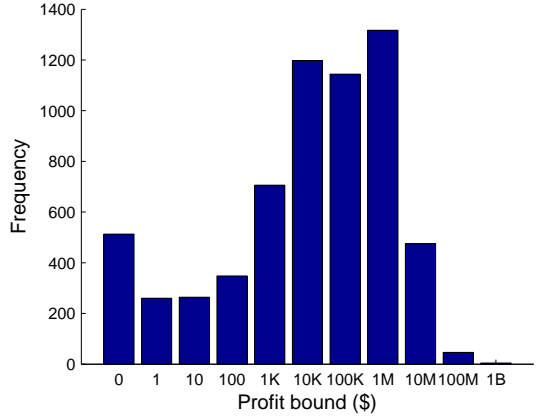}
\caption[]{\small Histogram of profit bounds for US stocks (10 second
  holding)}
\label{fig:profithist}
\end{figure}

Figure~\ref{fig:regs} shows the fit between profits and the number of
TAQ quotes achieved by a simple two-parameter power law model for the
same 19 stocks.  The $R^2$ value is 0.968.  Using this regression, we
can quickly estimate potential profits for a full universe of 6,279 US
stocks, yielding a bound on the total HFT profits available in all of
2008 at 10 second holding: \$21.3 billion.
Figure~\ref{fig:profithist} shows a histogram of our profit bounds
across these stocks.  Note that in these analyses we have included
only equities, which are the primary domain of retail traders, and not
other exchange-traded instruments like futures or ETFs. Inclusion of
these would increase our profitability estimates but not significantly
alter our conclusions.

\begin{figure}[tbp]
\centering \includegraphics[width=2.8in]{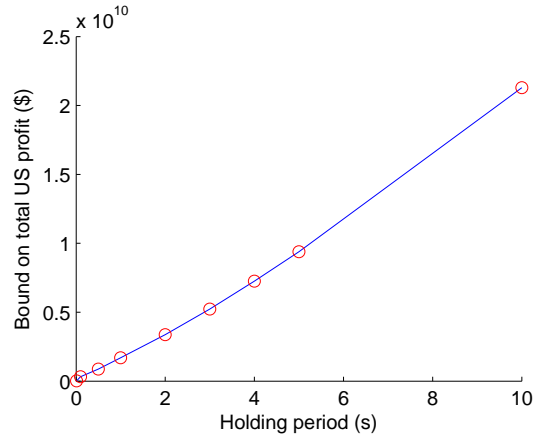}
\caption[]{\small Upper bound on profits for all US stocks vs holding
  period}
\label{fig:finalcurve}
\end{figure}

Finally, applying the same primary-to-composite and regression
methodologies to each holding period yields the curve shown in
Figure~\ref{fig:finalcurve}.  As discussed at the outset, profits fall
off sharply from a high of \$21 billion at 10-second holding to just
\$21 million at 10 milliseconds.  While these bounds are only
estimates --- the extrapolation process is necessarily imperfect ---
they give a rough picture of the profits we might expect from our OT,
which are themselves likely to be large overestimates of the profits
that could be realized in practice.

\section{Perspective}
\label{sec:perspective}

We conclude by offering some perspective on the size of the profit
bounds in Sections~\ref{sec:qat} and \ref{sec:taq}.  The total annual
trading volume in the US stock market (measured from TAQ data) is
approximately \$52 trillion, thus the maximum theoretical profit
reaped by aggressive HF traders on the US market ---
i.e., the largest number reported in this work --- is less than 0.05\%
of trading volume.

Furthermore, it is important to reiterate that we vastly overestimate,
in a number of ways, the profits that could actually be achieved by a
real-world trader:
\begin{itemize}
\item We assume no trading fees or commissions are paid by the HFT.
\item We assume the trader is omniscient and can exactly predict
  future price movements.
\item We assume the trader knows not only whether to buy or sell, but
  also precisely the optimal number of shares to trade at every
  moment.
\item We assume that these perfect predictions are made infinitely
  fast, and that trades execute with zero latency.
\item We assume that the trader's actions do not influence the market
  or create adverse price movement.
\item We assume that offers taken by the trader remain on the books,
  allowing the trader to repeatedly profit from a single opportunity
  as often as 100 times per second.
\end{itemize}

All of these assumptions fail in practice, reducing realizable profit.
For example, institutional traders might conservatively incur trading
fees of about 0.6 cents per share.  Since a large fraction of the
price fluctuations within short holding periods are very small, these
fees can be significant.  In order to achieve the \$3.4 billion
profits reported in Section~\ref{sec:qat}, the OT would need to trade
a total of 195 billion shares; if each such share cost 1.2 cents
(since it must be transacted twice to realize a profit), the total
trading fees would be \$2.3 billion, reducing profits by a full two
thirds.

Furthermore, the job of predicting short term price fluctuations
created by thousands of competing traders, each with a financial
incentive to remove any predictability, presumably cannot be performed
with anything like omniscient accuracy.  We believe that recognizing
even 10\% of the profitable opportunities is a phenomenally difficult
achievement in the real world.  Assuming this bar was not surpassed,
our results in Section~\ref{sec:taq} imply that 2008 HFT profits on
the entire US stock market were bounded by \$2.1 billion.  Of course,
a real trader will not only fail to act on some profitable
opportunities, but also mistakenly act in unprofitable cases, causing
additional losses.  This is especially true when considering trading
fees that must be paid regardless of whether a trade is profitable.

There are a number of caveats and limitations to our findings as well:
\begin{itemize}
\item We consider only a discrete set of fixed holding periods.  There
  may be additional profit opportunities that appear only at
  intermediate values of $h$.  However, preliminary experiments in
  which the fixed holding period for each trade is replaced with the
  most profitable holding period (up to $h$) for that trade result in
  only a modest increase in profits (less than 50\%).
\item We do not allow the OT to enter a postion with a market order but exit
with a limit order (or vice versa).
\item While large trades typically create adverse price movement, we do not
account for the possibility that a HF trader could profit by
manipulating the market in the opposite way, for example, by
generating excitement or panic.  Of course, if this were so, there
would be incentive for other HF traders to detect and revert such
manipulations on a similarly short timescale.
\item Finally, we again emphasize that our study is entirely limited to
aggressive HFT, and does not address the potential profitability of
passive, liquidity-providing HFT. Our justifications for this limitation
have been detailed.
\end{itemize}

In the end, we believe that our results demonstrate a surprisingly low
bound on the profitability of aggressive HFT.  While undoubtedly a
nontrivial source of income for some firms, profits from HFT, and
consequently the costs to traders without the technological
infrastructure necessary to compete, are relatively small.

We remark that 2008 was generally believed to be a banner year for HFT
profitability due to the volatility of markets during the financial
crisis.  Our preliminary experiments on 2009 data suggest that 2009
HFT profits were indeed about half of our 2008 estimates, although
this remains to be carefully verified.

\bibliographystyle{abbrv}
\bibliography{hft}

\end{document}